\documentstyle[prl,aps,multicol,psfig,epsfig]{revtex}

\begin{document}
\draft

\title{Memory Effects in Electron Transport in Si Inversion Layers in the
Dilute Regime: Individuality versus Universality}
\author{V.\ M.\ Pudalov$^{a,b}$, M.\ E.\ Gershenson$^{a}$, and H.\ Kojima$^{a}$}
\address{$^a$ Department of Physics and
Astronomy, Rutgers University, New Jersey 08854, USA}
\address{$^{b}$ P.\ N.\ Lebedev Physics
Institute, 119991 Moscow, Russia }

\maketitle

\begin{abstract}
In order to separate the universal and sample-specific effects in
the conductivity of high-mobility Si inversion layers, we studied
the electron transport in the same device after cooling it down to
4\,K at different fixed values of the gate voltage $V^{\rm cool}$.
Different $V^{\rm cool}$  did not modify significantly either the
momentum relaxation rate or the strength of electron-electron
interactions. However, the temperature dependences of the
resistance and the magnetoresistance in parallel magnetic fields,
measured in the vicinity of the metal-insulator transition in 2D,
carry a strong imprint of individuality of the quenched disorder
determined by $V^{\rm cool}$. This demonstrates that the observed
transition between ``metallic'' and insulating regimes involves
both universal  effects of  electron-electron interaction and
sample-specific effects. Far away from  the transition, at lower
carrier densities and lower resistivities $\rho < 0.1 h/e^2$, the
transport and magnetotransport become  nearly universal.
\end{abstract}
\begin{multicols}{2}

After almost a decade of intensive research, the apparent
metal-insulator transition (MIT) in two-dimensional (2D) systems
remains a rapidly evolving field \cite{krav_RMP}. The central
problem in this field is whether the anomalous low-temperature
behavior of the conductivity, observed in high-mobility structures
in the dilute limit, signifies a novel quantum ground state in
strongly-correlated systems, or this is a semiclassical effect of
disorder  on electron transport. Indeed,  a great body of
experimental data demonstrates that, at least at sufficiently
large carrier densities (consequently, weak interactions), the low-temperature
behavior of disordered systems is governed by the universal
quantum corrections to the conductivity \cite{aa}. On the other
hand, there are also observations that near the apparent 2D MIT,
the behavior of dilute systems is very rich, and does not
necessarily follow the same pattern (e. g., even for the same
system, as Si MOSFETs, the "critical" resistance and the
high-field magnetoresistance vary significantly for different
samples \cite{noscaling,aniso}). This duality
(universality versus individuality) is reflected in two approaches
to the theoretical description of the "metallic" regime: some
models, based on strong electron-electron interactions, treat the
transition as a universal phenomenon
\cite{finkelstein,dicastro,punnoose,das,dolgopolov_MR,aleiner},
whereas the others emphasize the role of a sample-specific
disorder (traps, localized spins, potential fluctuations, etc.)
\cite{am,amp,kozub,dasklap,meir}.

In order to find  a definite experimental answer to this problem
and to separate universal and non-universal (``individuality'')  effects in the
vicinity of the 2D MIT, we have studied
the electron
transport in the same Si MOS structure, which was slowly cooled
down from room temperature to $T=$4\,K at different fixed values
of the gate voltage $V_g=V^{\rm cool}$ \cite{cooldown}.
Changing the cooling conditions allowed us to vary the confining potential and
the density of quenched localized states without affecting the
main parameters which control electron-electron interactions in
the system of mobile electrons: the momentum relaxation rate and
the interaction constants. By tuning $V_g$ at
low temperatures, we varied the electron density $n$ over the
range $n=(0.7-3)\times10^{11}$cm$^{-2}$ in a system
with a snapshot disorder pattern. Two key features of the 2D MIT,
strong dependences of the resistivity on the temperature and
parallel magnetic field, have been studied.

We have observed two distinct regimes as a function of the
electron density. At relatively high densities
(resistivity $\rho \leq 0.1 h/e^2$), the dependences $R(T)$
and $R(B_\parallel)$ in weak parallel magnetic fields
$B_\parallel$ are similar for different cooldowns; the similarity
indicates `universal' behavior.
In contrast,
at low densities ($\rho \sim (0.1-1) h/e^2$), or in moderate and
strong parallel fields $g\mu_B B_\parallel \sim E_F \gg k_BT $,
the cooling conditions affect dramatically the
electron transport. This observation provides direct experimental
evidence that the behavior of dilute systems becomes
sample-specific near the apparent 2D MIT no matter how one
approaches the transition (either by decreasing the electron
density, or by increasing the parallel magnetic field).

The resistivity measurements were performed on a high mobility
Si-MOSFET sample \cite{sample} at the bath temperatures $0.05 -
1.2$\,K. The crossed magnetic field system allowed to accurately
align the magnetic field parallel to the plane of the 2DEG
\cite{gm}. The carrier density, found from the period of
Shubnikov-de Haas (SdH) oscillations, varies linearly with $V_g$:
$n \approx C\times (V_g-V_{\rm th})$, where $C$ ($=1.103\times
10^{11}$\,/Vcm$^2$ for the studied sample) is determined by the
oxide thickness. The  `threshold' voltage
$V_{\rm th}$ varied little (within 0.15\,V) for
different cool-downs and remained fixed as long as the sample was
maintained at low temperatures (up to a few months).

Figure 1 shows the mobility $\mu$ versus $V_g$ for five different
cool-downs  with $V^{\rm cool}=0, 5, 10, 18$,  and  25\,V. The
peak mobility for different cool-downs varies by less than $\sim
7\%$); this demonstrates that the momentum relaxation time $\tau$
is not strongly affected by the cooling conditions. We also
observed that the amplitudes of the SdH oscillations are similar
for different cool-downs, as shown in the inset to Fig.~1. These
two observations are  consistent with each other, since the
quantum lifetime $\tau_q$ is nearly equal to $\tau$ for
Si-MOSFETs.

\vspace{0.15in}
\begin{figure}
\centerline{\psfig{figure=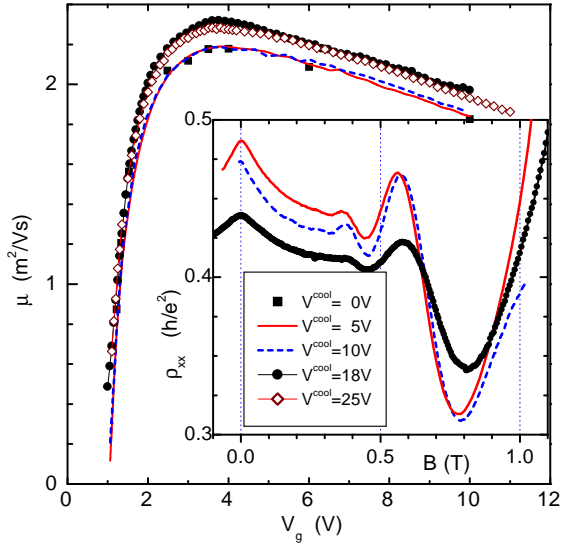,width=210pt}}
\vspace{0.1in}
\begin{minipage}{3.2in}
\caption{The mobility  versus the gate voltage for different
cool-downs. The $V^{\rm cool}$ values for both the main panel and
the inset are shown in the figure. Examples of the SdH
oscillations, shown in the inset for the same $V_g=1.15$\,V,
$T=0.1$\,K, $B_{\parallel}=0.03$\,T, demonstrate that the quantum
time $\tau_q$ is not very sensitive to the cooling conditions. The
carrier densities are (from top to bottom) $n=1.081$, $1.092$,
$1.070$ in units $10^{11}$cm$^{-2}$.} \label{fig1}
\end{minipage}
\end{figure}

Figure 2\,a shows the dependences $\rho(T)$ for two cool-downs in
the vicinity of the 2D MIT. Far from the transition, where
$\rho\ll h/e^2$, the dependences $\rho (T)$ are
cooldown-independent. However, in the vicinity of the transition
($\rho\sim h/e^2$), a dramatically different behavior is
observed. The irreproducibility of $\rho(T)$ for different
cool-downs is clearly seen for the curves in Fig.~2 which
correspond to the same $\rho$ at the lowest $T$: these curves,
being different at higher temperatures, converge with decreasing $T$.
The sample-specific memory effects vanish also at sufficiently low
temperatures: this suggests that the underlying mechanism is
related to the finite-temperature effects in a system which
retains a quenched disorder.
These results suggest that, in addition to universal effects,
a finite-temperature and sample-specific mechanism, which strongly
affects the resistivity, comes into play.

The `critical' density $n_c$, which corresponds to the transition,
was found from a linear extrapolation to zero of the density dependence
of the activation energy
$\Delta(n)$ in the insulating regime $\rho(T)\propto
\exp(\Delta/T)$ \cite{noscaling,akk}. The dependences $\rho(T)$, which
corresponded to $n=n_c$ for two cool-downs shown in Fig.~2a, are
highlighted in bold. It is clear that (i) the `critical' densities
and `critical' resistivities depend on the cooling conditions, and
(ii) the `critical' dependences $\rho(T, n=n_c)$ are non-monotonic
(see also Refs.~\cite{akk,reznikov}). Observation of the
non-monotonic critical dependences $\rho_c(T)$ suggests that a)
saturation of the temperature dependences at $n=n_c$, reported in
\cite{kk}, is not a universal effect, and b) the critical density
is not necessarily associated with the sign change of
$d\rho/dT(n)$ \cite{akk}.

\vspace{0.15in}
\begin{figure}
\centerline{\psfig{figure=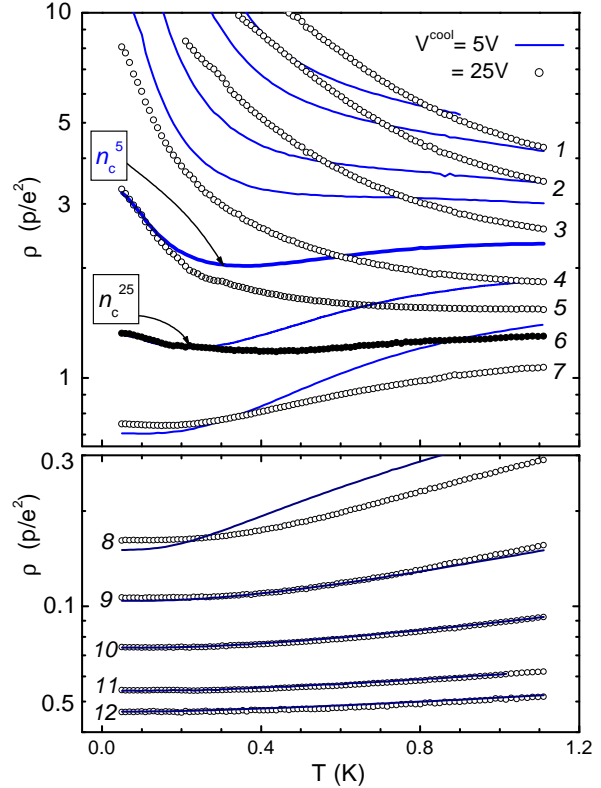,width=220pt} }
\vspace{0.1in}
\begin{minipage}{3.2in}
\caption{Temperature dependences of the resistivity for two
different cool-downs. The densities, which correspond to curves
{\it 1} to {\it 12}, are as follow: 0.783, 0.827, 0.882, 0.942,
0.972, 1.001, 1.021,  1.31, 1.53, 1.87, 2.29, 2.58 in unites
of $10^{11}$cm$^{-2}$.
}
 \label{fig2}
\end{minipage}
\end{figure}

We now turn to the
magnetoresistance (MR) in parallel fields; the
data are shown in Figs.~3 and 4. This MR is usually associated with the
spin effects \cite{krav_RMP,aniso}. In the theoretical models
of the parallel-field MR, based on electron-electron interactions,
the MR is controlled by the effective $g^*$-factor and the momentum
relaxation time  $\tau$ \cite{finkelstein,aleiner,lee&rama}. An
important advantage of our method is that cooling of the same
sample at different $V^{\rm cool}$ does not affect these parameters.
Thus, one might expect to observe a sample-independent behavior if
the MR is controlled solely by the universal interaction effects.

Firstly, let us consider the range of fields much weaker than the
field of complete spin polarization ($g^*\mu_B B_{\parallel}\ll
E_F$). The insets to Figs. 3\,a and 3\,b  show that the MR is
proportional to $B_{\parallel}^2$ at $g^*\mu_B
B_{\parallel}/k_BT\leq 1$. We found that the slope $d\rho/dB^2$ is
nearly cooldown-independent (i.e. universal) only for the densities $n >1.3\times
10^{11}$cm$^{-2}$ (which are by 30\% greater than the critical
density $n_c$), or for the resistivities  $\rho < 0.16h/e^2$
(compare insets to Figs.~3\,a and 3\,b). With approaching $n_c$,
this universality vanishes: Figure 3\,a shows that even when the
zero-field resistivity is as small as $0.22h/e^2$, the slope
varies by a factor of 1.3 for different $V^{\rm cool}$.

For the intermediate fields, $k_BT<g^*\mu_B B_{\parallel}<E_F$,
the $\rho(B_{\parallel})$ behavior is not universal over the whole
 density range $n=(1-3)\times 10^{11}$cm$^{-2}$ (Figs.~3a,b). As $n$
decreases and approaches $n_c$, the cooldown-dependent variations
of $R(B_{\parallel})$ increase progressively.
Observation of large ($\sim 50\%$) non-universal variations  of
the MR at intermediate fields makes the scaling analysis
\cite{shashkin} of the MR in this field range dubious.

\begin{figure}
\centerline{\psfig{figure=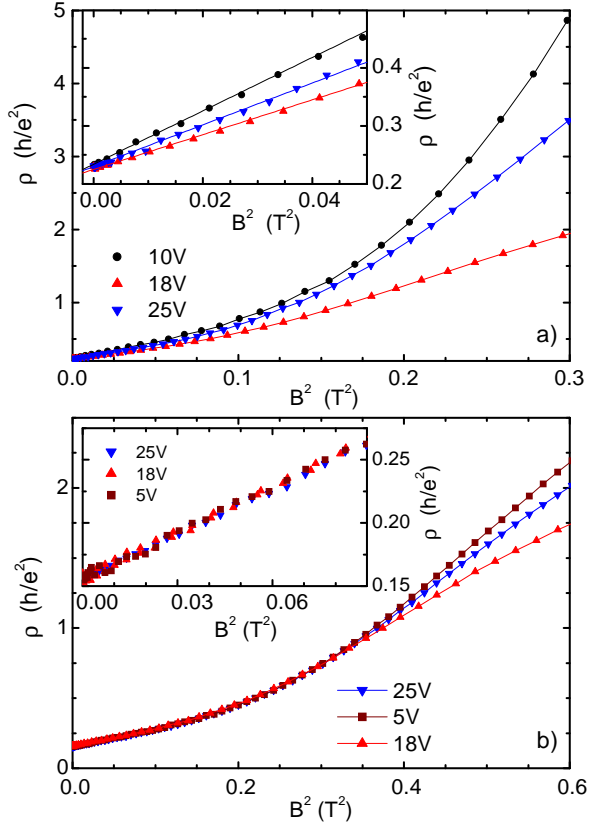,width=220pt} }
\vspace{0.1in}
\begin{minipage}{3.2in}
\caption{Examples of the dependences  $\rho (B_{\parallel}^2)$ at
$T=0.3$\,K for the carrier density (a) $1.20\times
10^{11}$\,cm$^{-2}$ and (b) $1.34\times 10^{11}$\,cm$^{-2}$. The
insets blow up the low-field region of the quadratic behavior. The
values of $V^{\rm cool}$ are indicated for each curve.}
\label{fig3}
\end{minipage}
\end{figure}

The influence of cooldown conditions becomes even more dramatic in
strong fields, $B \gtrsim E_F/g^*\mu_B$. Despite the fact that the
dependences $\mu (n)$ for different cool-downs are very similar
(Fig. 1), we observed very large variations in the strong-field
MR. Figures 4\,a and 4\,b show  $R(B_{\parallel})$ for different
cool-downs at two values of $n$. The cooldown conditions cause
factor-of-five changes in $\rho(B)$ in high fields  and
factor-of-two changes in the values of $B_{\parallel}=B_{\rm sat}$
at which the MR `saturates' at a given carrier density. The latter
quantity was determined from the intercept of the tangents at
fields below and above MR saturation \cite{aniso}.

Figure 4\,c shows the dependences $B_{\rm sat}(n)$  for different
cool-downs. We also plotted here the density dependence of the
field $B_{\rm pol}  = 2 E_F^*/g^*\mu_B = n\pi\hbar^2/m^*g^*\mu_B
$, for the complete spin polarization of mobile electrons ($m^*$
is the renormalized effective mass). The dependence $B_{\rm
pol}(n)$ was calculated using sample-independent (universal)
$g^*m^*$ values  \cite{gm}. Comparison between $B_{\rm sat}$ and
$B_{\rm pol}$ shows that $B_{\rm sat}$ does not necessarily
manifest the complete spin polarization, and that the coincidence
of $B_{\rm sat}$ with $B_{\rm pol}$  reported in
Ref.~\cite{vitkalov_doubling} may be rather accidental. This
non-universal, sample-dependent behavior of $B_{sat}$
agrees with earlier observations
made on different samples \cite{aniso}. We emphasize that the
curves for different $V^{\rm cool}$ (in each of Figs.~4\,a and b)
correspond to the same  density,
as follows from Hall voltage and/or SdH oscillations measurements.
The fact that $B_{\rm
sat}$ is a cooldown-dependent parameter, suggests that the MR in
strong parallel fields is not solely related to spin-polarization
of mobile electrons; we speculate that it also reflects the spin
polarization of the sample-specific localized electron states,
which might have the effective g-factor quite different from that
for the mobile electrons \cite{aniso}.

\begin{figure}
\centerline{\psfig{figure=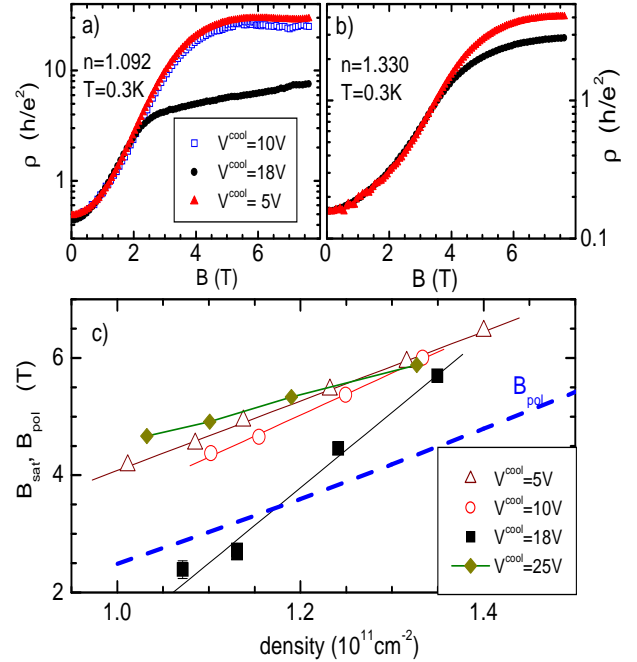,width=230pt,height=250pt} }
\vspace{0.1in}
\begin{minipage}{3.2in}
\caption{Resistivity vs in-plane magnetic field for three
cool-downs at two densities:
a) $n=1.092\times 10^{11}$cm$^{-2}$ and b) $n=1.33\times
10^{11}$cm$^{-2}$. c) The saturation field $B_{\rm sat}$ versus
$n$ for four different cool-downs. Solid lines are  guides to the
eye. Dashed line shows the field of complete spin polarization
calculated on the basis of direct measurements of the spin
susceptibility for mobile electrons \protect\cite{gm}. }
 \label{fig4}
\end{minipage}
\end{figure}

It is worth mentioning that the influence of variable disorder on
transport and magnetotransport in Si-MOSFETs has been observed
earlier. Both the temperature dependence $\rho(T)$
and magnetoresistance $\rho(B_\parallel)$ were found to be
different in samples with different mobility
\cite{noscaling,aniso} and in samples cooled down with different
values of substrate bias voltage \cite{pepper}. In these studies,
however, the sample mobility was changed significantly. In
contrast, in our studies we kept  constant the sample mobility
and all parameters relevant to electron-electron
interaction.

To summarize, by cooling the same high-mobility Si-MOS sample from
room temperature down to $T=4$K at different fixed values of the
gate voltage,
we tested universality of the temperature and
magnetic-field dependences of the resistivity near the
2D MIT. An important advantage of this approach is that the
different cooldown procedures do not affect the parameters which
control the contribution of the interaction effects to the
resistivity. It has been
found that in the vicinity of the transition ($\rho \sim h/e^2$),
the sample-specific effects strongly affect $\rho(T)$; these
effects vanish only when $\rho$ decreases below $\sim 0.1h/e^2$
with increasing electron density.
 Non-universal behavior has been also observed
for the  magnetoresistance in parallel magnetic fields. The effect
is especially dramatic in moderate ($E_F^*> g^*\mu_B B
>T$) and strong ($g^*\mu_B B \gtrsim E_F^*$) fields: it extends to
much higher electron densities (we observed pronounced
non-universality of $R(B_{\parallel})$ over a wide range
$n=(1-3)\times 10^{11}$cm$^{-2}$). Our results clearly demonstrate
that the apparent metal-insulator transition in 2D involves both
universal and sample-specific effects. These results also help to
establish the borderline between the
regimes where the electron transport in high-mobility Si MOS-structures is
either universal or sample-specific.
Understanding of the non-universal regime requires
detailed knowledge of the interface
disorder at low temperatures.

Authors acknowledge fruitful discussions with E.\ Abrahams, B.\ L.\ Altshuler
and D.\ L.\ Maslov.

The work was partially supported by the NSF, INTAS, NATO, NWO, and
Russian Programs RFBR, `Statistical physics', `Integration', and
`The State support of the leading scientific schools'.


\end{multicols}


\begin{references}
\bibitem{cooldown}
$V^{\rm cool}$ determines the depth of the confining potential well
and, simultaneously, the number of interface
traps sunk under
the Fermi level, and the density of localized states
from the tails of upper subbands.
At low temperatures, whenever gate voltage is varied,
the potential well remains almost unchanged
and memorizes an imprint  of the disorder formed during its cooling down.
\bibitem{sample} Sample Si6-14 with 190\,nm thick gate oxide
was fabricated on  (001)-Si wafer and had  source-drain
orientation along [100]. For more details, see Ref.~\protect\cite{gm}.
\end{references}
\end{document}